\documentclass[conference]{IEEEtran}
\ifCLASSINFOpdf
\else
\fi
\usepackage{amsmath}
\usepackage{amssymb}
\usepackage{amsthm}
\usepackage{algorithm}
\usepackage{algpseudocode}
\usepackage{graphicx}
\usepackage{subfig}
\usepackage{cite}

\theoremstyle{definition}

\newtheorem{theorem}{Theorem}
\newtheorem{definition}{Definition}
\newtheorem{lemma}{Lemma}

\makeatletter
\renewcommand{\maketag@@@}[1]{\hbox{\m@th\normalsize\normalfont#1}}
\makeatother

\allowdisplaybreaks[4] %

\begin{document}
%
\title{\huge{A Cascaded Channel-Power Allocation for D2D Underlaid Cellular Networks Using Matching Theory}}

\author{
\IEEEauthorblockN{Yiling Yuan\IEEEauthorrefmark{1}, Tao Yang\IEEEauthorrefmark{1}, Yuedong Xu\IEEEauthorrefmark{1}, Hui Feng\IEEEauthorrefmark{1} and Bo Hu\IEEEauthorrefmark{1}\IEEEauthorrefmark{2}}
\IEEEauthorblockA{
\IEEEauthorrefmark{1} Research Center of Smart Networks and Systems, School of Information Science and Engineering\\
\IEEEauthorrefmark{2}Key Laboratory of EMW Information (MoE)\\
Fudan University, Shanghai, China, 200433\\
Emails: \{yilingyuan13, taoyang, ydxu, hfeng, bohu\}@fudan.edu.cn}}

\maketitle

\begin{abstract}
We consider a device-to-device (D2D) underlaid cellular network, where each cellular channel can be shared by several D2D pairs and only one channel can be allocated to each D2D pair. We try to maximize the sum rate of D2D pairs while limiting the interference to cellular links. Due to the lack of global information in large scale networks, resource allocation is hard to be implemented in a centralized way. Therefore, we design a novel distributed resource allocation scheme which is based on local information and requires little coordination and communication between D2D pairs. Specifically, we decompose the original problem into two cascaded subproblems, namely channel allocation and power control. The cascaded structure of our scheme enables us to cope with them respectively. Then a two-stage algorithm is proposed. In the first stage, we model the channel allocation problem as a many-to-one matching with externalities and try to find a strongly swap-stable matching. In the second stage, we adopt a pricing mechanism and develop an iterative two-step algorithm to solve the power control problem.
\end{abstract}


%
\IEEEpeerreviewmaketitle

\section{Introduction}
Device-to-device (D2D) communication as an underlay to cellular networks is one of the key technologies to meet the dramatically increasing traffic demand and provide better user experience in future cellular networks\cite{Asadi2014A}. The basic idea is to allow nearby mobile devices to reuse cellular spectrum by establishing direct communication links without interacting with base station (BS).
\par
One big challenge for implementing underlaid D2D communication is how to allocate spectrum resource efficiently. To date, numerous resource allocation schemes have been proposed \cite{Feng2013Device,Zhao2015Resource,Hoang2015Resource}. In \cite{Feng2013Device}, an efficient scheme was developed to jointly optimize the channel allocation and power control. Nevertheless, only one D2D pair was allowed to use one cellular channel, which may limit the system throughput. The spectrum efficiency of the cellular system can be improved further if multiple D2D pairs are allowed to share the same channels\cite{Zhao2015Resource,Hoang2015Resource}.
\par
Most of existing resource allocation schemes worked in centralized way. These schemes are developed under the assumption that the channel state information (CSI) between every transmitter and receiver is available to a central controller (e.g., the BS), which incurs heavy overhead. Therefore, it is more preferable to design a distributed resource allocation scheme with limited local channel information\cite{Maghsudi2015Hybrid,Ye2015Distributed,Yin2013Distributed,Hasan2015Distributed,Nguyen2016Distributed}. In \cite{Maghsudi2015Hybrid}, authors studied the system that allowed the channels to be shared by several D2D pairs. Nonetheless, the proposed algorithm lacked an efficient distributed power control approach to guarantee the service level of cellular user (CU). In \cite{Ye2015Distributed, Yin2013Distributed,Hasan2015Distributed,Nguyen2016Distributed}, authors investigated the system where D2D pairs could reuse all the channels. However, the D2D pairs in close proximity will suffer from severe mutual interference, which may make the interference management complicated.
\par
In this paper, we consider the system where each channel can be shared by multiple D2D pairs and each D2D pairs can reuse at most one channel at each slot. The distributed resource allocation schemes for such system are less explored in the literature. Unlike \cite{Maghsudi2015Hybrid},  we try to maximize the sum rate of D2D pairs while guaranteeing the service level of CUs. We decompose the original problem into two cascaded subproblems: channel allocation and power control, and then a two-stage distributed algorithm is proposed.
\par
Specifically, in the first stage, the channel allocation problem is modelled as many-to-one matching, which is suitable for assignment problem between two disjoint sets of players with local information. Unlike many existing works for resource allocation, the proposed matching game has externalities, which are resulted from the mutual interference among D2D pairs sharing the same channel. A distributed algorithm is proposed to find a strongly swap-stable matching as solution. Moreover, the existence of strongly swap-stable matching is proved. In the second stage, a pricing mechanism is adopted and an iterative two-step algorithm is presented to solve the power control problem. At the beginning, a virtual price factor based on the received interference is broadcast. Then, each D2D pair independently maximizes its utility according to the virtual price factor. The virtual price factor acts as control signal to limit the interference. Our contribution is that the proposed scheme can be implemented via distributed decision at each device based on local information. Moreover, the cascaded structure can reduce the exchange of control signal in need. The numerical simulations show the proposed algorithm with limited CSI is efficient and the throughput loss compared to brute-force method is small in our setup.
\par
The rest of this paper is organized as follows. In Section II, the system model and problem formulation are established. In Section III, a two-stage algorithm is proposed to solve the two subproblems respectively. The Section IV gives the numerical results. Finally Section V concludes this paper.

\section{System Model and problem formulation}
\subsection{System Model}
We study a D2D underlaid cellular network comprised of a BS, $C$ CUs and $D$ D2D pairs. The scenario where D2D pairs reuse uplink resource is considered in this paper. The set of CUs and D2D pairs are denoted by $\mathcal{C}=\{1,\cdots,C\}$ and $\mathcal{D}=\{1,\cdots,D\}$ respectively. All devices are equipped with one antenna. The network is provided with a set $\mathcal{K}$ of $K$ orthogonal frequency channels. We assume a fully loaded cellular network, i.e., $K=C$. Each channels has been already allocated to one cellular user. For simplicity, CU $k\in\mathcal{K}$ is referred to the CU assigned to channel $k$ in our discussion. Multiple D2D pairs can share the same cellular channels and each D2D pair is allowed to access at most one cellular channel. The set of D2D pairs sharing the channel $k$ is denoted by $\mathcal{D}_k\subseteq\mathcal{D}$, and $\mathcal{D}_k\cap\mathcal{D}_k'=\emptyset$ when $k\neq k'$. Then, the SINR of D2D pair $i$ on channel $k$ is given by
\begin{equation}\label{equ1}
  \gamma^D_{dk} = \frac{p_d g^k_{dd}}{n_0+q_c g^k_{d} + \sum_{i\in\mathcal{D}_k\setminus\{d\}}p_i g^k_{id}},
\end{equation}
where $q_c$ and $p_d$ are the transmit powers of cellular link and D2D pair $d$, respectively, $g^k_{id}$ denotes the channel gain from D2D transmitter $i$ to D2D receiver $d$ on channel $k$, $g^k_{d}$ is the channel gain from CU $k$ to D2D receiver $i$ on channel $k$, and $n_0$ is the noise power.
\subsection{Problem Formulation}
In this paper, we aim to maximize the sum rate of D2D pairs while guaranteeing the maximum interference to cellular users. Mathematically, the problem can be formulated as
\begin{subequations}
\label{equ2}
  \begin{align}
      \max_{\mathbf{p},\mathbf{D}}  & \ \ \sum_{k\in\mathcal{K}}\sum_{d\in\mathcal{D}_k} \text{ln}(1+\gamma^D_{dk})\\
      \rm{s.t.}                 & \ \ 0 \leq p_d \leq P_m, \forall d\in\mathcal{D}, \label{equ2:1}\\
                                & \ \ \sum_{d\in\mathcal{D}_k}p_d h^k_d \leq Q_k, \forall k\in\mathcal{K}, \label{equ2:2}\\
                                & \ \ \mathcal{D}_k\cap\mathcal{D}_k'=\emptyset,\forall k,k'\in\mathcal{K},k\neq k', \label{equ2:3}
  \end{align}
\end{subequations}
where $\mathbf{p} = (p_1,p_2,\cdots,p_D)^T$, $\mathbf{D} = \{\mathcal{D}_1,\mathcal{D}_2,\cdots,\mathcal{D}_K\}$, $P_m$ is the maximum transmit power and $h^k_d$ denotes the channel gain from D2D transmitter $d$ to cellular link on channel $k$. The constraint (\ref{equ2:2}), referred to as \emph{interference constraint}, is for protection of cellular links, where $Q_k$ is the interference tolerance level which depends on the requirements and channel gain of cellular link on channel $k$. Moreover, we assume only local information is available at each device. Explicitly, D2D pair $d$ only knows the channel gain $h^k_d, g^k_{dd}, g^k_{d}, g^k_{id}$ and $g^k_{di}$, and the BS only knows $h^k_d$.\footnote{We assume the system works in TDD mode. Hence, due to channel reciprocity, these CSI can be easily obtained by listening to the pilot transmitted by CU or other devices .}
\begin{figure}[!t]
\centering
\includegraphics[width=3.1 in]{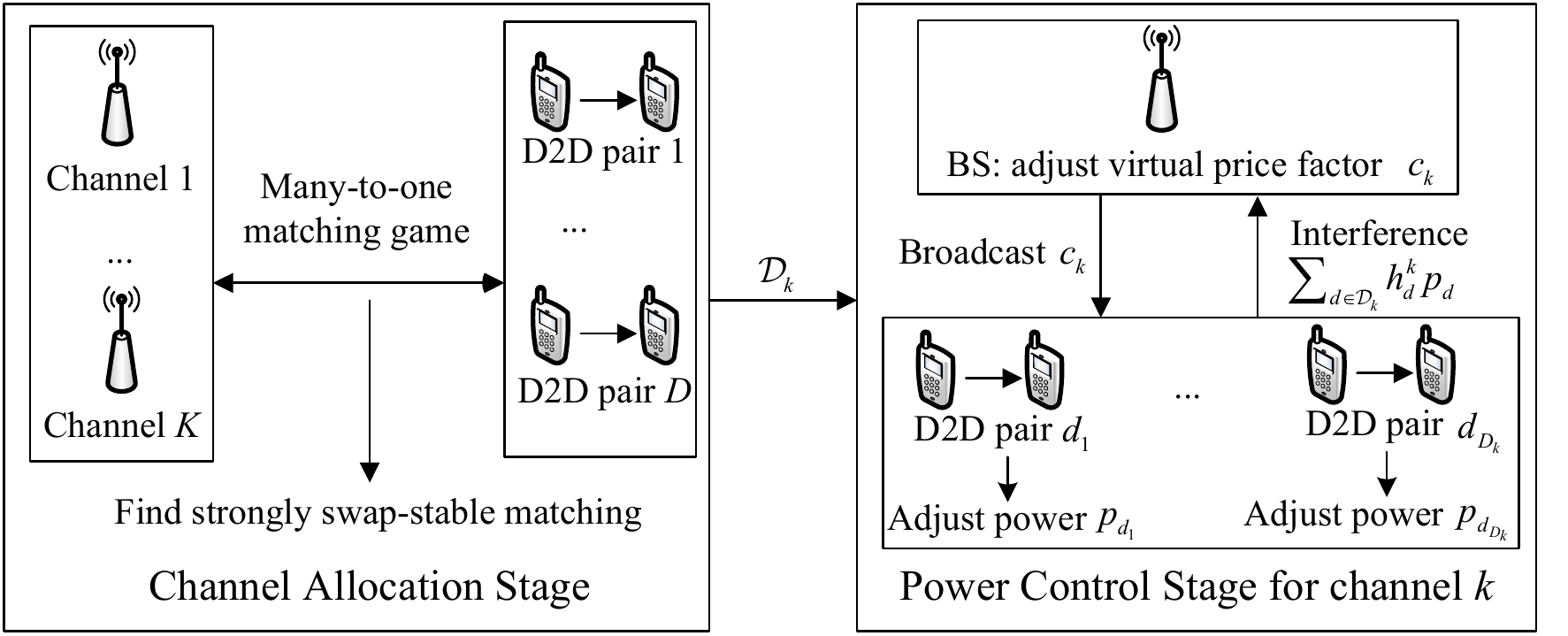}
\caption{Framework for resource allocation}
\label{framework}
\end{figure}
\section{Channel allocation and power control}
Problem (\ref{equ2}) is a mixed integer non-linear programming problem, which is usually intractable. Moreover, only local information is available, which makes the problem more difficult to solve. Therefore, our goal is to develop a efficient distributed approach. For this purpose, we decompose the problem into two cascaded subproblems, namely channel allocation and power control. The former one is solved by matching theory and a pricing mechanism is adopted for the latter one. The entire framework is depicted in Fig.\ref{framework}.

\subsection{Channel Allocation Stage}
We consider the optimization problem of the channel allocation solution given the transmit power $\mathbf{p}$ by solving the following optimization problem:
\begin{subequations}
\label{equ3}
  \begin{align}
      \max_{\mathbf{D}}  & \ \ \sum_{k\in\mathcal{K}}\sum_{d\in\mathcal{D}_k} \text{ln}(1+\gamma^D_{dk})\\
      \rm{s.t.}                 & \ \ \mathcal{D}_k\cap\mathcal{D}_k'=\emptyset,\forall k,k'\in\mathcal{K},k\neq k', \label{equ3:1}\\
                                & \ \ p_d = P_m, \forall d\in\mathcal{D}. \label{equ3:2}
  \end{align}
\end{subequations}
In problem (\ref{equ3}), the interference constraint (\ref{equ2:2}) will be considered in power control stage. In order to estimate the contribution of each D2D pair to the overall throughput, D2D transmitters are requested to transmit at power $P_m$, as represented in constraint (\ref{equ3:2}).
\par
To solve problem (\ref{equ3}), which is a combinational optimization problem, we model it as many-to-one matching game. Originally stemming from economics\cite{Roth1990Two}, matching theory provides a mathematical framework to cope with the problem of matching players in two distinct sets, based on each player's individual preference and local information. It has been used for many resource allocation problems in communication networks\cite{Gu2014Matching}. In our context, one side is D2D pairs and the other side is the cellular channels. In the implementation, the BS will make decisions on behalf of channel $k$.
\begin{definition}
\label{def:matching}
A \textbf{many-to-one matching} $\mu$ is a subset of $\mathcal{D}\bigotimes\mathcal{K}$ such that $|\mu(d)| = 1,|\mu(k)|=n_k,\forall d\in\mathcal{D},\forall k\in\mathcal{K}$, where $n_k$ is the quota of CU $k$, $\mu(d)=\{k\in\mathcal{K}:(d,k)\in\mu\}$ and $\mu(k)=\{d\in\mathcal{D}:(d,k)\in\mu\}$.
\end{definition}
If the number of D2D pair in $\mu(k)$, say $r$, is less than $n_k$, then $\mu(k)$ has $n_k-r$ ``holes'' represented as D2D pairs with no preference over matchings. In this paper, we assume $n_k = D$ for simplicity. We will also use $\mu(d)$ to denote the channel reused by D2D pair $d$. Utility function is adopted to describe the preferences of agents over matchings. Given matching $\mu$, the utility of D2D pair $d$ is defined as follows:
\begin{equation}
\label{equ4}
U_d(\mu)  = \xi_1\phi_d(\mu) - \theta,
\end{equation}
where $\phi_d(\mu)$ is the gain obtained from accessing channel $\mu(d)$, $\xi_1$ is the equivalent revenue with respective to gain $\phi_d(\mu)$, and $\theta$ is the price for using cellular channel. We set $U_d(\mu)=-\infty$ when D2D pair $d$ is not matched with any channel. Specifically, the gain is represented as follows:
\begin{equation}
\label{equ5}
\begin{split}
\phi_d(\mu)  = &\ln(\frac{P_mg^{\mu(d)}_{dd}}{n_0})-wq_cg^{\mu(d)}_{d}\\
            &-\sum_{i\in\mu^2(d),i\neq d}\frac{w}{2}P_m(g^{\mu(d)}_{di} + g^{\mu(d)}_{id}),
\end{split}
\end{equation}
where $w$ is tradeoff coefficient and $\mu^2(d)$ denotes the set of D2D pairs reusing channel $\mu(d)$. The difference between the first two terms of (\ref{equ5}) can be regarded as the benefit from channel, which takes into consideration both the channel gain and interference from cellular link, and only depends on the matched channel itself. The last term is the loss resulted from the mutual interference between D2D pairs. The following lemma implies that $\sum_{d\in\mathcal{D}}\phi_d(\mu)$ can be regarded as a lower bound of the aggregated throughput of D2D pairs.
\begin{lemma}
Suppose that each D2D pair is matched with one channel. Then, given $p_d=P_m,\forall d\in\mathcal{D}$, for any matching $\mu$,
\begin{equation*}
\sum_{d\in\mathcal{D}}\phi_d(\mu) + \text{const.}<\sum_{k\in\mathcal{K}}\sum_{d\in\mu(k)} \text{ln}(1+\gamma^D_{dk}).
\end{equation*}

\end{lemma}
\begin{IEEEproof}
\begin{footnotesize}
\begin{align*}
        & \sum_{k\in{\mathcal{K}}}\sum_{d\in\mu(k)} \ln(1+\gamma^D_{dk})\\
   > & \sum_{k\in{\mathcal{K}}}\sum_{d\in\mu(k)} \ln(\frac{P_m g^k_{dd}}{n_0+q_c g^k_{d} + \sum_{i\in\mu(k)\setminus\{d\}}P_m g^k_{id}})\\
    =   & \sum_{k\in{\mathcal{K}}}\sum_{d\in\mu(k)} \Bigg\{\ln({wP_m g^k_{dd}}) - \ln \Big[w(n_0+q_c g^k_{d} + \sum_{\substack{i\in\mu(k)\\ i\neq d}}P_m g^k_{id})\Big]\Bigg\} \\
   \overset{(a)}{\geq} & \sum_{k\in{\mathcal{K}}}\sum_{d\in\mu(k)} \Bigg\{\ln({\frac{P_m g^k_{dd}}{n_0}}) -  w(q_c g^k_{d} + \sum_{i\in\mu(k)\setminus\{d\}}P_m g^k_{id})\Bigg\} + \text{const.}\\
    =  &\sum_{d\in\mathcal{D}}\phi_d(\mu) + \text{const.},
\end{align*}
\end{footnotesize}
where the inequality (a) follows from the standard logarithm inequality, $\ln{x}\leq x-1,\forall x>0$.
\end{IEEEproof}
\par
In addition, the utility of channel $k$ is defined as follows:
\begin{equation}
\label{equ6}
U_k(\mu) = \theta|\mu(k)|- \xi_2C(\sum_{d\in\mu(k)}\frac{P_m h^k_d}{Q_k} - 1),
\end{equation}
where $C(x)=\max(0,x)$ quantifies the degree of violation of the constraint (\ref{equ2:2}) and $\xi_2$ is the cost coefficient.
Note that (\ref{equ4}) and (\ref{equ6}) can be calculated locally at each device.
\par
Because of the interference between D2D pairs, the utility of D2D pair $d$ will be affected by the choice of other D2D pairs. Thus the proposed matching has \emph{externalities} and is called many-to-one matching with externalities\cite{Roth1990Two}. Due to the \emph{externalities}, the stable matching, which is a solution concept used widely in matching theory, may not exist. To this end, we look at another stability concept, based on the concept of swap-matching\cite{Bodine2011Peer}:
\begin{definition}
\label{def:swap-mathing}
Given a matching $\mu$, two D2D pairs $s,t\in\mathcal{D}$ and two channels $m,n\in\mathcal{K}$ with $(s,m),(t,n)\in\mu$, a \textbf{swap-matching} is $\mu^s_t=\mu\setminus\{(s,m),(t,n)\}\cup\{(s,n),(t,m)\}$.
\end{definition}
In the swap-matching, two involved D2D pairs exchange their matched channels while all other matchings remain the same. Moreover, one of the D2D pair can be a ``hole'', thus allowing D2D pairs to move into available vacancies.
\begin{definition}
\label{def:swap-stable}
  A matching $\mu$ is \textbf{strongly swap-stable} if and only if there exists no swap-matching $\mu^s_t$ such that: $U_s(\mu^s_t)\geq U_s(\mu),U_t(\mu^s_t)\geq U_t(\mu)$ and $U_{\mu(s)}(\mu^s_t)+U_{\mu(t)}(\mu^s_t)\geq U_{\mu(s)}(\mu)+U_{\mu(t)}(\mu)$, and at least one of the above three inequalities is strict.
\end{definition}

 Note that the above swap-stability is stronger than that defined in \cite{Bodine2011Peer}. The rationality of our stability notion comes from the observation that the BS aims to maximize the total utility of CUs and can permit monetary transfer among CUs. Algorithm \ref{alg:channel allocation} is proposed to find a strongly swap-stable matching. The initialization step is based on Gale-Shapely (GS) algorithm. Each D2D pair $d$ ranks channels based on the descending order of estimated SINR. Meanwhile, each channel $k$ ranks the D2D pair according to the ascending order of interference channel gain $h^k_{d}$. Then, each D2D pair proposes to its most preferred channel, and each channel accepts the most preferred $n_k$ D2D pair and rejects the others. Initialization step terminates when each D2D pair is accepted by a channel, which can be guaranteed by the assumption $n_k=D$. Next, the algorithm seeks ''approved'' swap-matching  until no such swap-matching exists. 
\begin{algorithm}[!t]
\caption{ Stage 1: Channel Allocation for D2D Pairs}
\label{alg:channel allocation}
\begin{small}
\begin{algorithmic}[1]
\renewcommand{\algorithmicrequire}{\textbf{GS-algorithm-based initialization:}}
\Require
\State D2D pairs and channels construct their preference lists based on estimated SINR $\tilde{\gamma}^k_d=\frac{P_{m}g^k_{dd}}{n_0+q_cg^k_{d}}$ and interference channel gain $h^k_{d}$, respectively.
\State Each D2D pair proposes to its most preferred channel that has not rejected it before.
\State Each channel keeps the most preferred $n_k$ D2D pairs and rejects the others.
\State Repeat step 2 and 3 until each D2D pair is accepted by a channel.
\renewcommand{\algorithmicrequire}{\textbf{Swap operation:}}
\Require
\State Each D2D pair $s$ searches for swap-matching $\mu^s_t$ ''approved'' by the players involved in the swap. Then, $\mu=\mu^s_t$.
\State If $\mu$ is changed, repeat step 5.
\end{algorithmic}
\end{small}
\end{algorithm}
\par
To prove the convergence of Algorithm \ref{alg:channel allocation}, we introduce a potential function for the proposed game:
{
\begingroup\makeatletter\def\f@size{9}\check@mathfonts
\def\maketag@@@#1{\hbox{\m@th\large\normalfont#1}}%
\begin{equation*}
\label{equ7}
\begin{split}
  \Phi(\mu) = &\sum_{k\in{\mathcal{K}}}\sum_{d\in\mu(k)} \xi_1\Big\{\ln(\frac{P_m g^k_{dd}}{n_0}) - w(q_c g^k_{d} + \sum_{\substack{i\in\mu(k)\\ i\neq d}}\frac{P_m g^k_{id}}{2})\Big\} \\
              &- \sum_{k\in{\mathcal{K}}}\xi_2C(\sum_{d\in\mu(k)}\frac{P_m h^k_d}{Q_k} - 1).
\end{split}
\end{equation*}
\endgroup}
\begin{theorem}
Algorithm \ref{alg:channel allocation} always converges to a strongly swap-stable matching.
\end{theorem}
\begin{IEEEproof}
The proof is based on  the fact that  the potential function is improved. after each swap operation in Algorithm 1, which is proved as follows.
\par
Note that each D2D pair is matched with one channel after initialization. Assume the ''approved'' swap-matching is $\mu^s_t$. For convenience, let $W_d(\mu) = \xi_1\ln(\frac{P_mg^{\mu(d)}_{dd}}{n_0})-\xi_1wq_cg^{\mu(d)}_{d}-\theta$. For $i,j\in\mathcal{D},k\in\mathcal{K}$, $I^k(i,i) = 0$ and $I^k(i,j) = \xi_2wP_m(g^k_{ij}+g^k_{ji})/2$. Moreover, we assume $\mu(s)=m,\mu(t)=n$. Thus
{
\begingroup\makeatletter\def\f@size{9}\check@mathfonts
\def\maketag@@@#1{\hbox{\m@th\large\normalfont#1}}%
\begin{equation*}
  \label{equ8}
  \begin{split}
        & \Phi(\mu^s_t)-\Phi(\mu) \\
      = & \sum_{k\in\mathcal{K}}U_k(\mu^s_t) + \sum_{d\in\mathcal{D}}W_d(\mu^s_t)-\frac{1}{2}\sum_{k\in\mathcal{K}}\sum_{i,j\in\mu^s_t(k)}I^k(i,j)\\
        & -\left\{\sum_{k\in\mathcal{K}}U_k(\mu) + \sum_{d\in\mathcal{D}}W_d(\mu)-\frac{1}{2}\sum_{k\in\mathcal{K}}\sum_{i,j\in\mu(k)}I^k(i,j)\right\}.
  \end{split}
\end{equation*}
\endgroup}
Expanding and canceling the like terms, and using the symmetric property of $I(i,j)$, we obtain
{
\begingroup\makeatletter\def\f@size{9}\check@mathfonts
\def\maketag@@@#1{\hbox{\m@th\large\normalfont#1}}%
\begin{equation*}
\label{equ9}
  \begin{split}
        & \Phi(\mu^s_t)-\Phi(\mu) \\
      = & U_m(\mu^s_t)-U_m(\mu) + U_n(\mu^s_t)-U_n(\mu)\\
        &+ W_s(\mu^s_t)-W_s(\mu) + W_t(\mu^s_t)-W_t(\mu)\\
        &+ \sum_{i\in\mu(m)} I^m(i,s) - \sum_{i\in\mu(m)} I^m(i,t) + I^m(s,t)\\
        &+ \sum_{i\in\mu(n)} I^n(i,t) - \sum_{i\in\mu(n)} I^n(i,s) + I^n(s,t).
  \end{split}
\end{equation*}
\endgroup}
Considering the utilities of players involved in the swap, we can find out that
\begin{equation*}\label{equ10}
\begin{split}
   \Phi(\mu^s_t)-\Phi(\mu)  = & U_m(\mu^s_t) + U_n(\mu^s_t) + U_s(\mu^s_t) + U_t(\mu^s_t)\\
                              &- \{U_m(\mu) + U_n(\mu) + U_s(\mu) + U_t(\mu)\}.
\end{split}
\end{equation*}
Thus we can conclude that the potential function is improved ager swap. The proof is similar when one of D2D pairs is a "hole".
\par
Furthermore, he number of possible matching between D2D pairs and channels is limited, Algorithm 1 will terminate at finite steps. On the other hand, the algorithm does not terminate until there is no any "approved" swap-matching. Therefore, we can conclude that the final matching is swap-stable.
\end{IEEEproof}

\subsection{Power Control Stage}
Since all the cellular channels are orthogonal, we can decouple the power control problem into $K$ subproblems, where we consider each channel independently. For simplicity, we focus on the power control problem of channel $k$. Similar to \cite{Ye2015Distributed, Maghsudi2015Hybrid}, we adopt a pricing mechanism consisted of two steps. In the first step, the BS determines the virtual price factor $c_k$ to control the received interference. In the second step, with the virtual price information, each D2D pair adjusts transmit power aiming to maximize its utility. To solve the two-step problem, the backward induction technique is adopted. We start with the transmit power determination problem, called lower problem. Then we investigate the price factor adjustment problem, called upper problem.
\subsubsection{Lower Problem}
Because each D2D pair $d\in\mathcal{D}_k$ maximize its own utility independently with local information, it is natural to model the lower problem as non-cooperative game. The power control game model is defined as $\mathcal{G}_k=\{\mathcal{D}_k,\{\mathcal{A}_d\}_{d\in\mathcal{D}_k},\{R_d\}_{d\in\mathcal{D}_k}\}$, where $\mathcal{D}_k = \{d_1,\cdots,d_{D_k}\}$ is the set of $D_k$ D2D pairs assigned to channel $k$, $\mathcal{A}_d = [0,P_m]$ is the set of joint action profiles of all players, and $R_d:\mathcal{I}\rightarrow \Re^+$ is the payoff function of player $d$. The payoff function is defined as follows:
\begin{equation}
\label{equ11}
  R_d(\mathbf{p}_k) = \ln(\gamma^D_{dk}(\mathbf{p}_k)) - c_kh^k_dp_d,
\end{equation}
where $\gamma^D_{dk}(\mathbf{p}_k)$ is defined in (\ref{equ1}) and $\mathbf{p}_k = (p_{d_1},\cdots,p_{d_{D_k}})^T$ is the action profile of all players. The first term can be considered as reward, which is an approximation of achievable rate. The second term is the cost charged by the BS, which is proportional to the interference caused by this D2D pair to cellular link.
\par
We will adopt a well-studied solution notion known as Nash Equilibrium (NE)\cite{Osborne1994A}, from which no players has intention to unilaterally deviate. A NE of our game model is given in Theorem 2.
\begin{theorem}
 Given $c_k$, $\mathbf{p}^*_k = (p^*_{d_1},\cdots,p^*_{d_{D_k}})^T$ is a NE of proposed power control game, where
 \begin{equation}
 \label{equ12}
   p^*_d = \min(P_m,\frac{1}{c_kh^k_d}),\forall d\in \mathcal{D}_k.
 \end{equation}
\end{theorem}

\begin{IEEEproof}
Let $\mathbf{p}_{-d}$ denote the joint action profile of all players except player $d$. Given $c_k$ and $\mathbf{p}_{-d}$, player $d$ would like to maximize its utility as follows:
\begin{subequations}
\label{equ13}
  \begin{align}
      \max_{p_d}  & \ \ \ln(\gamma^D_{dk}(p_d,\mathbf{p}_{-d})) - c_kh^k_dp_d\\
      \rm{s.t.}                 & \ \ 0 \leq p_d \leq P_m,
  \end{align}
\end{subequations}
The objective function is concave in $p_d$. After solving problem (\ref{equ13}), we can obtain the best response of player $d$: $BR_d(\mathbf{p}_{-d})=\min(P_m,\frac{1}{c_kh^k_d})$. Note that this best response implies that a rational player will always take a fixed action no matter what actions are taken by its opponents. Consequently, we can conclude that in NE, each player $d\in\mathcal{D}_k$ would take the action $p^*_d = \min(P_m,\frac{1}{c_kh^k_d})$.
\end{IEEEproof}

\subsubsection{Upper Problem}
In upper problem, the BS will adjust the virtual price factor $c_k$ to limit the interference to cellular link. On the one hand, too small $c_k$ will result in the interference exceeding the interference tolerance level. On the other hand, too large $c_k$ will result in low transmit power of D2D pairs, which leads to low sum rate of D2D pairs and inefficient spectrum utilization. Therefore,  it is natural to seek a appropriate $c_k$ which could maximize the sum rate of D2D pairs while guaranteeing the service level of CU. However, only interference information is available at the BS. Thus, it is difficult to find such $c_k$. To this end, instead of adjusting $c_k$ to obtain the optimal power allocation, we try to seek $c_k$ such that the power profile obtained in the following lower problem is Pareto optimal, which is defined as follows:
\begin{definition}
  A power profile $\mathbf{p}_k=(p_{d_1},\cdots,p_{d_{D_k}})$ satisfying constraint (\ref{equ2:2}) is \textbf{Pareto optimal} if there exists no power profile $\mathbf{p}_k$ satisfying constraint (\ref{equ2:2}) could improve one D2D pair's rate without deteriorating other D2D pairs' rates.
\end{definition}
\begin{theorem}
\label{Pareto Optimal}
If a power profile $\mathbf{p}_k$ satisfying $\sum_{d\in\mathcal{D}_k}p_d h^k_d = Q_k$, then it is Pareto optimal.
\end{theorem}
\begin{IEEEproof}
We will prove the theorem by contradiction. Suppose there is another power profile $\mathbf{p}'_k$ which can increase or maintain the performance of all D2D pairs. Then there must exist a set of D2D pairs $\mathcal{M}\subset\mathcal{D}_k$ such that for any D2D pair $d\in\mathcal{M}$, $p_d$ decreases. Let $m = arg\min_{d\in\mathcal{M}}\frac{p'_d}{p_d}$ and $\alpha = \frac{p'_m}{p_m}<1$. Consequently, $p'_d\geq\alpha p_d$ for $\forall d\in \mathcal{D}_k$. Hence
\begin{align*}
\gamma^D_{mk}(\mathbf{p}'_k)
              & \leq \frac{\alpha p_m g^k_{mm}}{n_0+q_c g^k_{m} + \alpha\sum_{i\in\mathcal{D}_k\setminus\{d\}}p_i g^k_{im}}\\
              & < \frac{ p_i g^k_{mm}}{n_0+q_c g^k_{m} + \sum_{i\in\mathcal{D}_k\setminus\{d\}}p_i g^k_{im}}\\
              & = \gamma^D_{mk}(\mathbf{p}_k),
\end{align*}
which is incompatible with our assumptions.
\end{IEEEproof}
Theorem \ref{Pareto Optimal} implies that the virtual price factor $c_k$ is ``appropriate'' if it can lead to a power profile which makes the interference constraint (\ref{equ2:2}) tight. Additionally, from (\ref{equ12}), it is easy to find out that $p^*_d$ is a non-increasing function of $c_k$. So a simple bisection algorithm can be used to find the ``appropriate'' $c_k$ according to the received interference at the BS. The algorithm solving the power control problem is depicted in Algorithm 2, where $c_{max}$ is the price such that $\sum_{d\in\mathcal{D}_k}p^*_d(c_{max}) h^k_d < Q_k$. According to Theorem \ref{Pareto Optimal} and the monotonicity of $p^*_d(c)$, Algorithm 2 will converge to the price factor leading to a Pareto optimal power profile. We consider non-trivial case, where the interference at the BS exceed $Q_k$ if all D2D pairs transmit using $P_m$, otherwise we just set $c_k = 0$.

\begin{algorithm}[!t]
\caption{ Stage 2: Power Control for channel $k$}
\label{alg:power control}
\begin{small}
\begin{algorithmic}[1]
\State Initialization: given channel $k$,$\mathcal{D}_k$, let $c_u=c_{max}$, $c_l = 0$;
\While {$|c_u-c_l|\geq\epsilon$}
\State The BS calculates $c_k = (c_u+c_l)/2$ and broadcasts it;
\State Each D2D pair $d\in\mathcal{D}_k$, calculate $p^*_d$ according to (\ref{equ12});
\If {$\sum_{d\in\mathcal{D}_k}p^*_d h^k_d < Q_k$}
\State $c_u = c_k$;
\Else
\State $c_l = c_k$;
\EndIf
\EndWhile
\State $c^*_k = c_k$;
\end{algorithmic}
\end{small}
\end{algorithm}

\section{Numerical Simulations}
The performance of our proposed algorithms is investigated through simulation in this section. The channel used in simulation is $h = \beta L^{-\eta}$, where $\beta$ is fast fading gain with exponential distribution, $\eta$ is the pathloss exponent and $L$ is the distance between transmitter and receiver. The D2D pairs and CUs are distributed uniformly within the cell. We set $\theta=\xi_1=\xi_2=1$ and $w=6\times10^6$. Other configuration parameters are depicted in Table.\ref{sim config}.
\begin{table}
\centering
\caption{Simulation Configure Parameters}
\label{sim config}
\begin{tabular}{|c|c|}
  \hline
  \textbf{Parameters}           &\textbf{Value}\\
  \hline
  Cell radius                   & 500 m\\
  Noise power($n_0$)            & -100 dBm\\
  Pathloss exponent($\eta$)     & 4\\
  Transmit power of CU($q_c$)               & 20 mW\\
  Maximum D2D power($P_m$)      & 20 mW\\
  Length of D2D links           & 50 m\\
  \hline
\end{tabular}
\end{table}

\begin{figure}[!t]
\centering
\subfloat[Performance of Algorithm 2 with $K = 1, D = 6$.]{\includegraphics[width=1.75in]{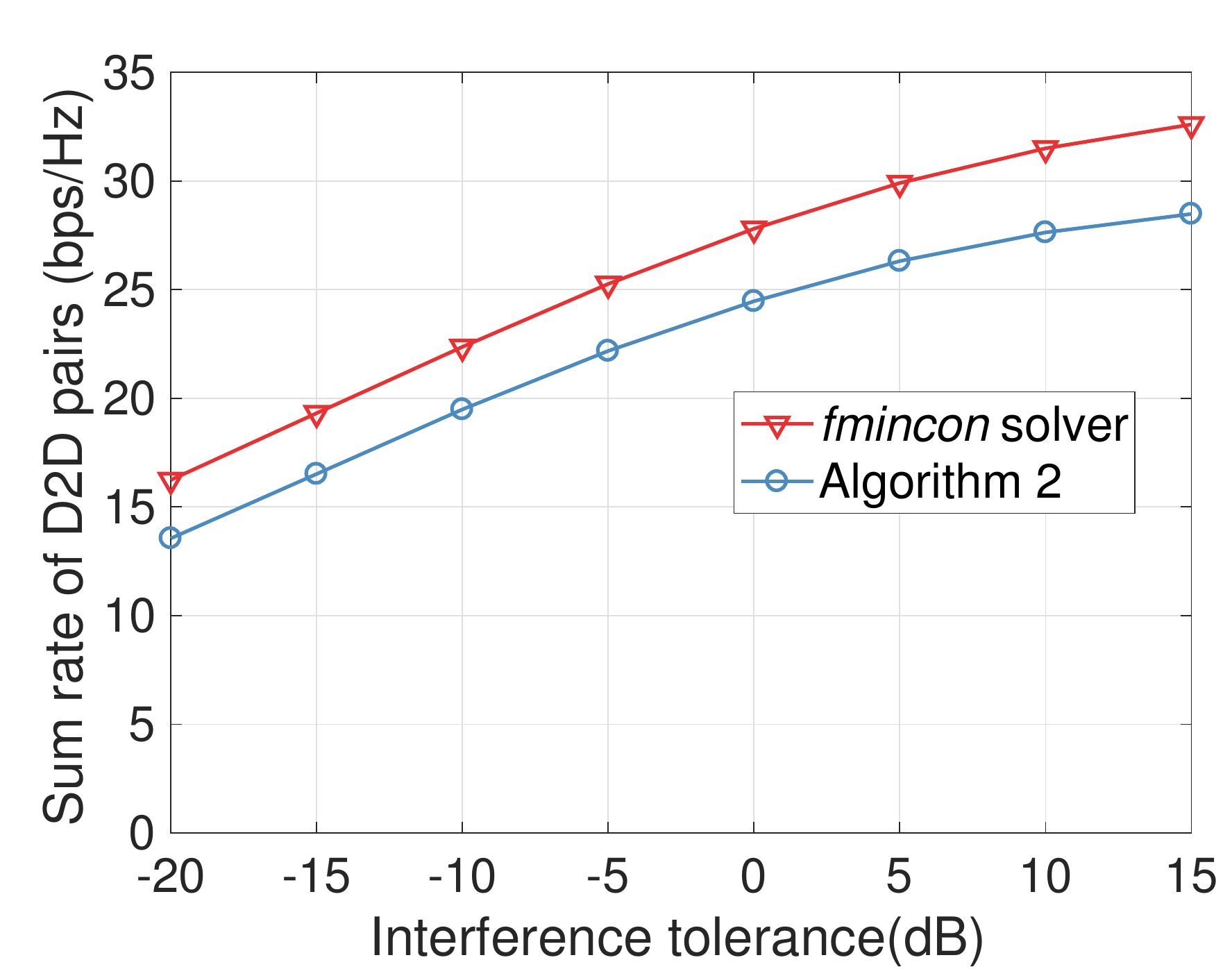}%
\label{alg2}}
\subfloat[Overall performance of our scheme with $K = 2, D = 6$.]{\includegraphics[width=1.75in]{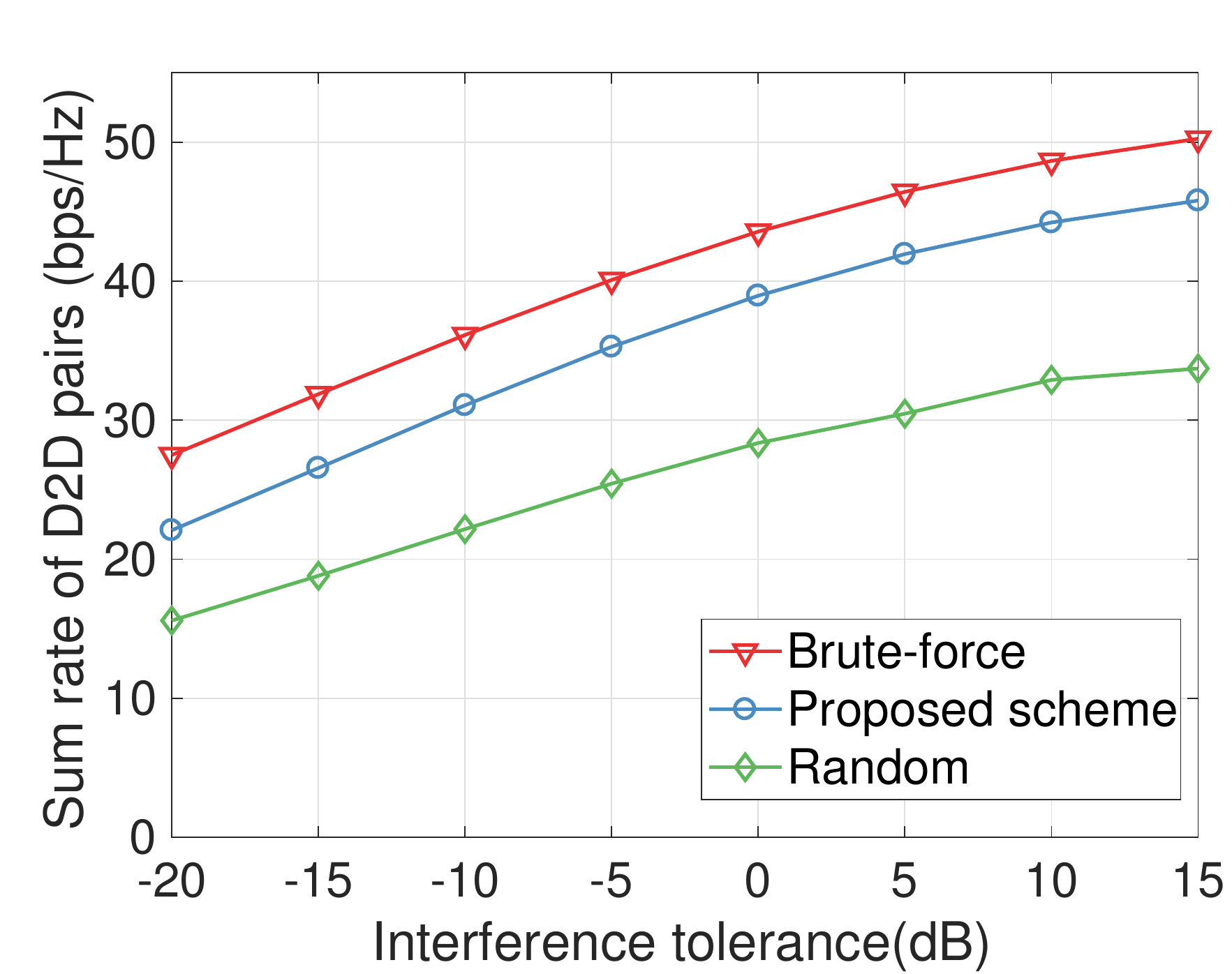}%
\label{total}}
\caption{Performance evaluation of our proposed scheme with different interference tolerance level.}
\label{PerformanceEvaluation}
\end{figure}
At first, we evaluate the performance of our proposed scheme with different interference tolerance levels normalized by the signal strength of cellular link. Fig.\ref{alg2} evaluates the achieved sum rate of D2D pairs of Algorithm 2. The results of locally optimal solution, which are obtained by \emph{fmincon} solver of \emph{Matlab}, are given for comparison. It can be seen that the performance of Algorithm 2 is close to the solution obtained by \emph{fmincon} solver. Fig.\ref{total} evaluates the overall performance of our scheme. The \emph{brute-force} scheme given for comparison is the algorithm which explores the all possible channel assignments and adopts \emph{fmincon} solver for power control. We can find that the performance loss of our scheme compared with \emph{brute-force} scheme is at most $20\%$ and is below $10\%$ when  interference tolerance level is larger than -5 dB, and the performance gain over \emph{random} scheme is larger than $35\%$.

\begin{figure}[!t]
\centering
\subfloat[Sum rate of D2D pairs.]{\includegraphics[width=1.75in]{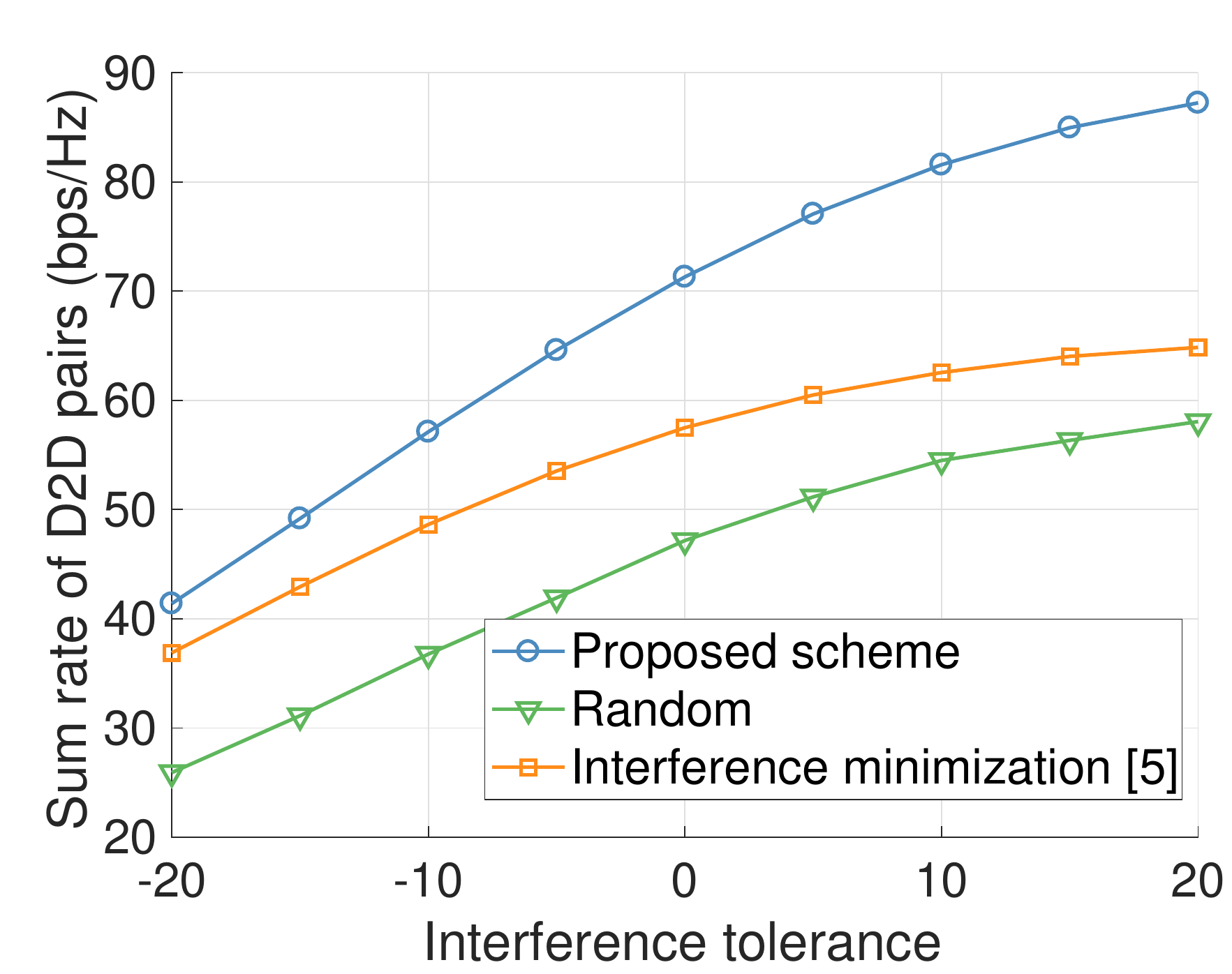}%
\label{CompareToleranceD2D}}
\subfloat[Sum rate of CUs.]{\includegraphics[width=1.75in]{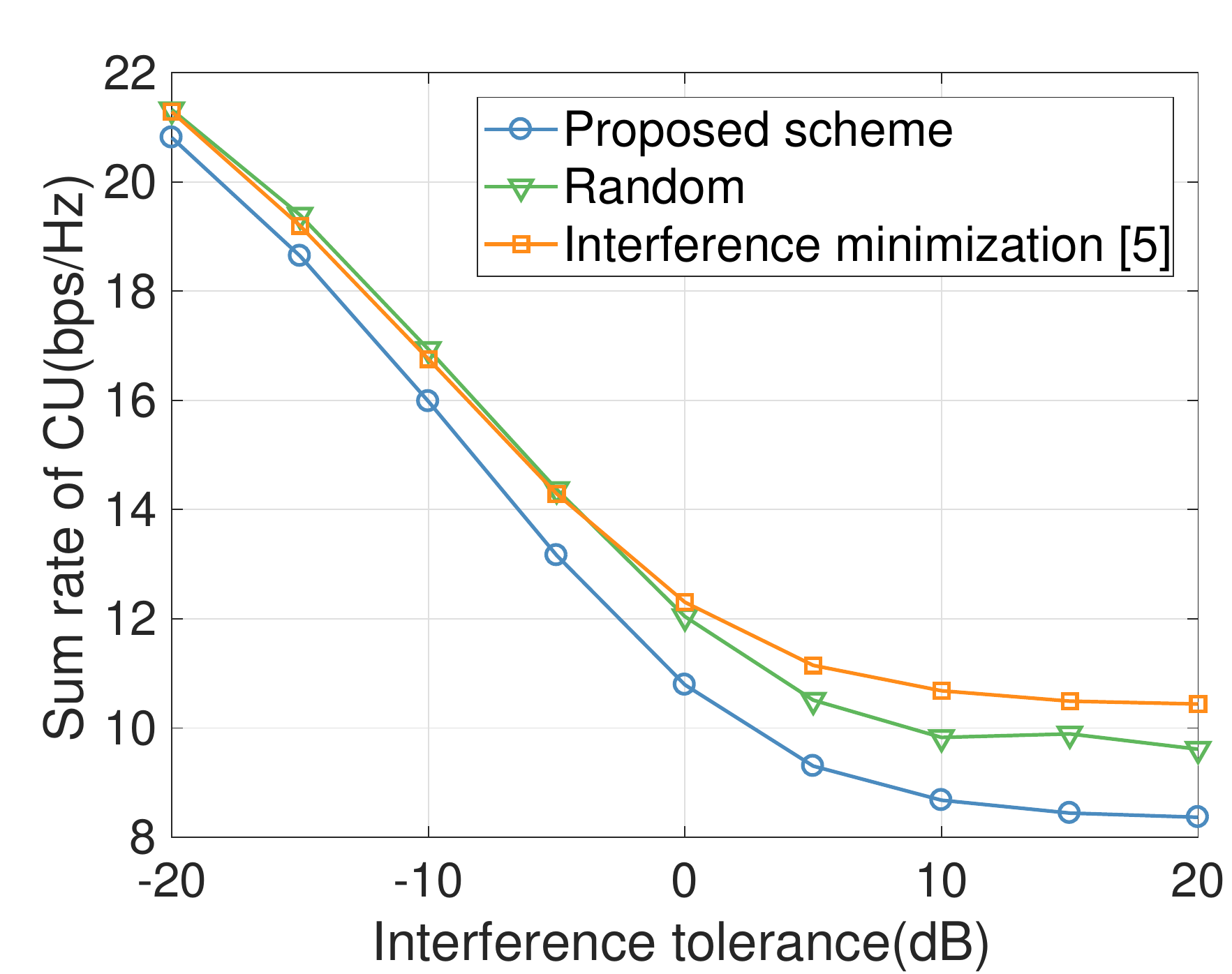}%
\label{CompareToleranceCU}}
\caption{Performance of different schemes with different interference tolerance level, $K = 4, D = 10$.}
\label{CompareTolerance}
\end{figure}

Next, Fig.\ref{CompareTolerance} compares the performance of different schemes for different interference tolerance level normalized by the signal strength of cellular link. The \emph{interference minimization} is referred to as the scheme in which $\sum_{k\in\mathcal{K}}\sum_{d\in\mathcal{D}_k}h^k_dP_m$ is minimized in channel allocation stage \cite{Maghsudi2015Hybrid} and Algorithm \ref{alg:power control}\\is used for power control stage. Fig.\ref{CompareTolerance} shows that as the interference tolerance level $Q_k$ increases, the sum rate of CUs decreases because D2D pairs are allowed to share the channel more aggressively which leads to more interference. On the other hand, the sum rate of D2D pairs increases as $Q_k$ increases, since D2D pairs are allowed to transmit at higher power. Moreover, adopting the proposed scheme, the D2D pairs can get much better performance, at the cost of less sum rate of CUs. However, such trade-off is worthy in terms of increasing spectrum efficiency. For instance, when $Q_k = 0\text{ dB}$, adopting our scheme, the performance loss of CU is about $12\%$ compared with other two schemes, but the sum rate of D2D pairs is at least $25\%$ greater than that of other schemes. Because of high-throughput of D2D pairs, the proposed scheme can yield considerable gain over other scheme in terms of the sum rate of entire system.

\begin{figure}[!t]
\centering
\includegraphics[width=3in]{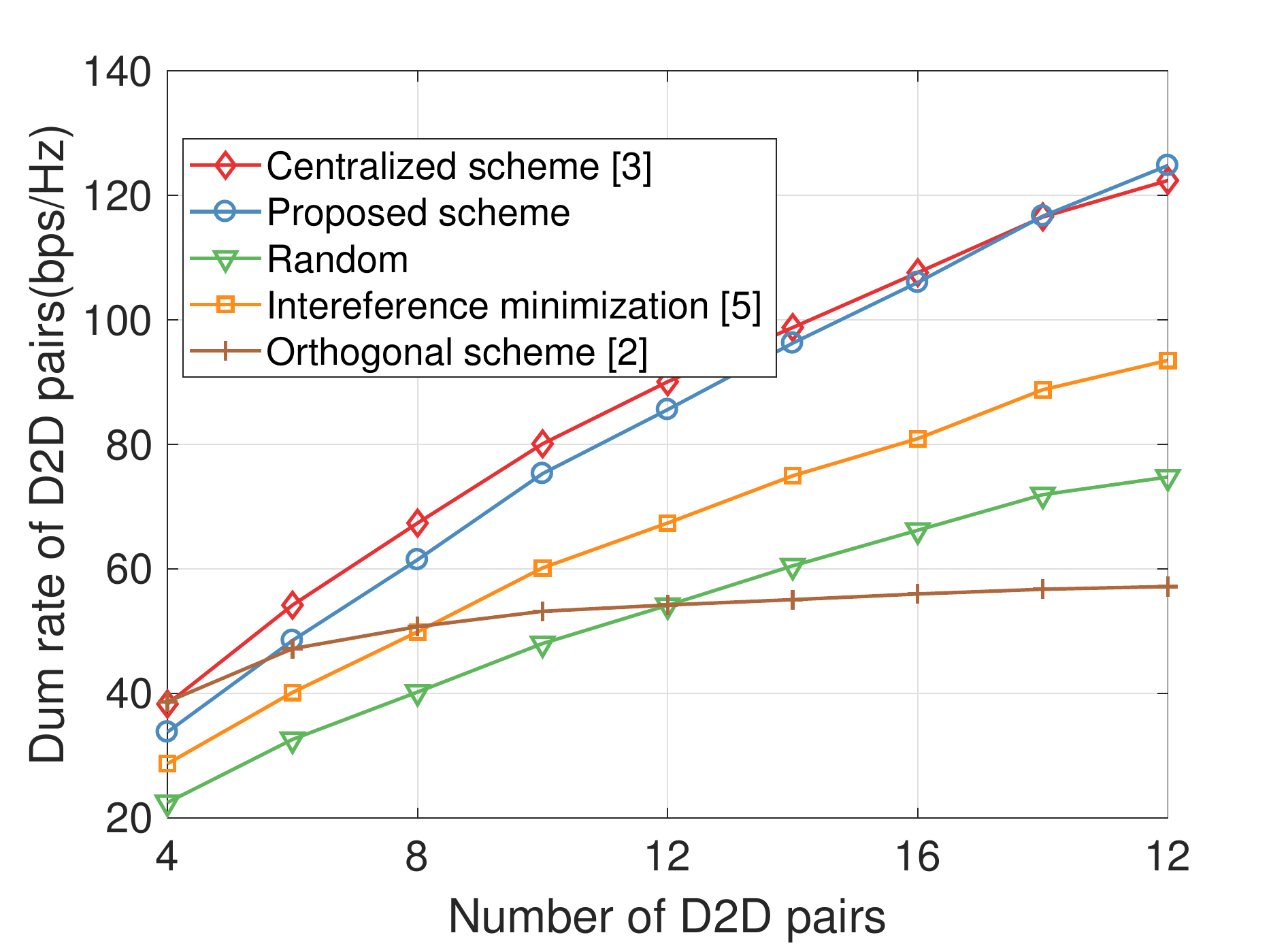}
\caption{The sum rate of D2D pairs with different number of D2D pairs, $Q_k=0\text{ dB},K=4$.}
\label{CompareD2DNum}
\end{figure}


Finally, Fig.\ref{CompareD2DNum} compares the sum rate of D2D pairs achieved by different schemes for different number of D2D pairs. We consider two another schemes as comparison, referred to as \emph{centralized} scheme and \emph{orthogonal} scheme respectively. The first scheme is proposed in \cite{Zhao2015Resource} to maximize the sum rate of D2D pairs in centralized way. The second scheme is similar to the algorithm proposed in \cite{Feng2013Device}, which assumes each channel is used by at most one D2D pair. This scheme chooses $K(K\geq D)$ D2D pairs to maximize the sum rate of D2D pairs. From Fig.\ref{CompareD2DNum}, we can see that the sum rate of the \emph{orthogonal} scheme grows approximately sublinearly as the number of D2D pairs increases, while the other three schemes grows almost linearly, which indicates that allowing multiple D2D pairs to reuse one cellular channel can improve the performance of entire system greatly. Moreover, the proposed scheme performs almost the same as the \emph{centralized} scheme and outperforms the other three schemes significantly. So we can conclude that our scheme is an efficient distributed resource allocation scheme in large scale D2D underlaid cellular networks.

\section{Conclusion}
In this paper, we propose a distributed resource allocation scheme for D2D underlaid cellular networks where each channel can be shared by one CU and several D2D pairs. We try to maximize the sum rate of D2D pairs while limiting the interference to cellular links. To solve the problem, we decompose the problem into two cascaded subproblems: channel allocation and power control problem, and a two-stage algorithm is proposed. Firstly, we model the channel allocation problem as a many-to-one matching game with externalities and try to find a strongly swap-stable matching. Secondly, we adopt price mechanism and propose an iterative two-step algorithm to solve the problem. Finally we present numerical results to verify the efficiency of our scheme. However, the interference information between D2D pairs is not fully utilized. In future work, such information can be used to design a more efficient algorithm for power distribution.
\section*{Acknowledgment}
This work was supported by the NSF of China (No. 61501124, No.71731004).

\bibliographystyle{IEEEtran}
\bibliography{IEEEabrv,reference}
\end{document}